\documentclass[pre,aps,twocolumn,showpacs]{revtex4}

\usepackage{graphicx}
\usepackage{amsmath}
\usepackage{amsfonts}
\usepackage{amssymb}
\usepackage[sort&compress]{natbib}

\usepackage{epstopdf}
\newcommand{\beq}{\begin{eqnarray}}
\newcommand{\eeq}{\end{eqnarray}}

\begin{document}

\title {Capillary-Gravity Waves Generated  by a Sudden Object Motion}
\author{F. Closa,$^{1}$ A.D. Chepelianskii,$^{2}$ and E. Rapha\"{e}l$^{1,\ast}$}
\affiliation{$^{1}$Laboratoire Physico-Chimie Th\'{e}orique, UMR CNRS GULLIVER 7083, ESPCI, 
10 rue Vauquelin, 75005 Paris, France}
\affiliation{$^{2}$Laboratoire de Physique des Solides, UMR CNRS 8502, B\^{a}timent 510, 
Universit\'{e} Paris-Sud, 91405 Orsay, France}
\date{\today}
\begin{abstract}
We study theoretically the capillary-gravity waves created at the water-air interface by a small 
object during a sudden accelerated or decelerated rectilinear motion.
We analyze the wave resistance corresponding to the transient wave pattern and show that it is 
nonzero even if the 
involved velocity (the final one in the accelerated case, the initial one in the decelerated case) is 
smaller than the minimum phase velocity $c_{min}=23 \mathrm{\,cm \, s^{-1}}$.
These results might be important for a better understanding of the propulsion of water-walking 
insects where accelerated and decelerated motions frequently occur.

\end{abstract}
\maketitle

\section{Introduction}

If a body (like a boat or an insect), or an external pressure source, moves at the free liquid-air 
interface, it generates \textit{capillary-gravity waves}. These are driven by a balance between the 
liquid inertia and its tendancy, under the action of gravity and under surface tension forces, to 
return to a state of stable equilibrium \cite{Landau}.  For an inviscid liquid of infinite depth, the 
dispersion relation of capillary-gravity waves, relating the angular frequency $\omega$ to the 
wavenumber $k$ is given by $\omega^{2}=gk+\gamma k^{3}/\rho$, where $\gamma$ is the liquid-air 
surface tension, $\rho$ the liquid density, and $g$ the acceleration due to gravity \cite
{Acheson}. The energy carried away by the waves is felt by the body (or the pressure source) as a 
drag $\mathrm{R}_{w}$, called the \textit{wave resistance} \cite{Lighthill}.
In the case of boats or ships, the wave resistance has been well studied in order to design hulls 
minimizing it \cite{Milgram}. The case of objects that are small compared to the capillary length $
\kappa^{-1}=\sqrt{\gamma/(\rho g)}$ has been considered only recently 
\citep{Elie:96,Elie:99,Keller,Chevy,Bacri,Steinberg1,Steinberg2}. 

In the case of a disturbance moving at constant velocity $V=|\mathbf{V}|$ on a rectilinear trajectory, 
the wave resistance $\mathrm{R}_{w}$ is zero for $V < c_{min}$ where $c_{min}=(4g\rho/
\gamma)^{1/4}$ is the minimum of the wave velocity $c(k)=\omega(k)/k=\sqrt{g/k+\gamma k/\rho}$ 
for capillarity gravity waves \citep{Lighthill,Lamb,Elie:96}. Effectively, in the frame moving with that object, the 
problem must be stationary. That is true only if the radiated waves have a phase velocity $c(k)$ 
equal to the object's velocity $V$. If $V<c_{min}$, no solutions exist, hence no waves and the wave resistance is zero.
For water with $\gamma = 73 \mathrm{ \, mN \, m^{-1}} $ and	$\rho = 10^{3} \mathrm{ \, kg \, m^
{-3}}$, one has $c_{min}=0.23 \mathrm{\, m \,s^{-1}}$ (at room temperature).
It was recently shown by Chepelianskii \textit{et al.} \cite{Alexei} that no such velocity threshold 
exists for a steady circular motion, for which, even for small velocities, a finite wave drag is 
experienced by the object. Here we consider the case of a sudden accelerated or decelerated 
rectilinear motion and show that the transient wave pattern leads to 
a nonzero wave resistance even if the involved velocity (the final one for the accelerated case, the 
initial one for the decelerated case) is smaller than the minimum phase velocity $c_{min}=23 
\mathrm{\,cm \, s^{-1}}$.
The physical origin of these results is similar to the Cherenkov radiation emitted by accelerated (or 
decelerated) charged particles \citep{Cherenkov,Jackson}.

\section{Equations of motion}\label{model} 

We consider an inviscid, deep liquid with an infinitely extending free surface. To locate a point on the free 
surface, we introduce a vector $\textbf{r}=(x,y)$ in the horizontal plane associated with the 
equilibrium state of a flat surface.
The motion of the disturbance in this plane induces a vertical displacement $\zeta(\mathbf{r},t)$ (Monge representation) of the free surface from 
its equilibrium position.

Assuming that the liquid equations of motion can be linearized (in the limit of small wave amplitudes), 
one has \cite{Alexei}

\begin{equation} \label{surface} 
 \frac{\partial ^{2} \hat {\zeta}(\mathbf{k},t)}{\partial t^{2}} \, + \, \omega(k)^{2}\hat {\zeta}(\mathbf{k},t) \; 
= \; -\frac{k \hat{P}_{ext}(\mathbf{k},t)}{\rho},
\end{equation} 
where $\hat{P}_{ext}(\mathbf{k},t)$ and $\hat{\zeta}(\mathbf{k},t)$ are the Fourier transforms of the 
pressure distribution and the displacement, respectively \cite{Fourier}. In what follows, we will 
assume that the pressure distribution is axisymmetric around the point $\mathbf{r}_{0}(t)$ 
(corresponding to the disturbance trajectory). $\hat{P}_{ext}(\mathbf{k},t)$ can then be written as $
\hat{P}_{ext}(k) e^{-i \mathbf{k} .\mathbf{r}_{0}(t)}$. 

\subsection{Uniform straight motion}

Let us first recall the results previously obtained in the case of a uniform straight motion \citep
{Lighthill, Lamb, Rayleigh, Kelvin}. Such a motion corresponds to $\mathbf{r}_{0}(t)=(-Vt,0)$ where 
$V$ is the constant velocity of the disturbance.

\begin{figure}
\includegraphics[width=3.2in,angle=00]{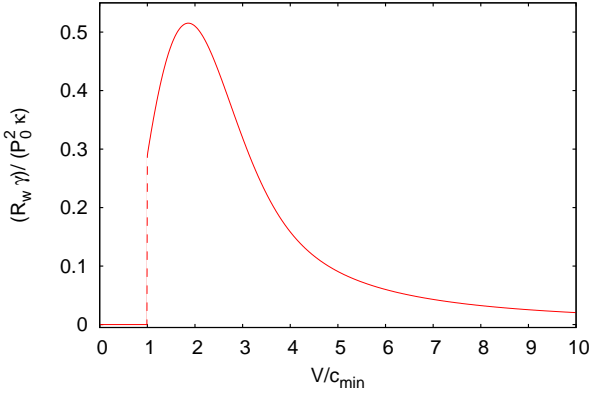}
\caption{Wave resistance $\mathrm{R}_{w}$ (in units of $P_{0}^{2}\kappa/\gamma$) as a function 
of the reduced velocity $V/c_{min}$ for a uniform straight motion, see Eq.(\ref{Rwave}). The 
pressure disturbance is assumed to be a Lorentzian, with a Fourier transform  $\hat P_{ext}(k)=P_
{0} e^{-bk}$, where b is the object size (set to $b=0.1\kappa^{-1}$).}
\label{fig1}
\end{figure}

Equation (\ref{surface}) then becomes
\begin{equation} \label{surface2} 
 \frac{\partial ^{2} \hat {\zeta }(\mathbf{k},t)}{\partial t^{2}} \, + \, \omega(k)^{2}\hat {\zeta}(\mathbf{k},t) 
\; = \; -\frac{k \hat{P}_{ext}(k) e^{i Vk_{x}t}}{\rho}.
\end{equation} 
The above equation corresponds to the equation of an  harmonic oscillator forced at angular 
frequency $Vk_{x}$. We can solve it by looking for solutions with a time-dependence of the form  
$e^{i Vk_{x}t}$, leading to
\begin{equation}\label{uniform} 
 \hat \zeta (\mathbf{k},t) \; = \; -\frac{k \hat{P}_{ext}(k)}{\rho (\omega(k)^{2}-(V k_{x})^{2})}e^{i Vk_{x}t}.
\end{equation} 
Following Havelock \cite{Lighthill}, the wave resistance $\mathbf{R}_{w}$ experienced by the 
moving disturbance is given by 

\begin{equation}\label{Hav} 
\mathbf{R}_{w}=-\mathit{i}\iint \! \dfrac{\mathrm{d} k_{x}}{2\pi} \, \dfrac{\mathrm{d} k_{y}}{2\pi} \, 
\mathbf{k} \; \hat \zeta (\mathbf{k},t) \, \hat P^{\ast}(k,t).
\end{equation} 
This expression for the wave resistance represents as the total force exerted by the external pressure on the free surface $\mathbf{R}_{w}=\iint \mathrm{d} x \mathrm{d} y P_{ext}(\mathbf{r},t)\triangledown \zeta(\mathbf{r},t)$  written in Fourier space.
Using Eq.(\ref{uniform}) and integrating over the angular variable one obtains \cite{Elie:96}

\begin{equation}\label{Rwave}
\mathbf{R}_{w} \; = \;  \displaystyle \int_{0}^{\infty}\!\! \dfrac{\mathrm{d}k}{2\pi} \, \frac{k P_{0}^{2}}
{\rho}\, e^{-2bk}\frac{\theta (V - c(k))}{V^{2}\sqrt{1-(c(k)/V)^{2}}} \;  \mathbf{u}_{x},
\end{equation} 
where $\theta (x)$ is the Heaviside step function and $\mathbf{u}_{x}$ the unit vector along the $x
$-axis.
The behavior of $\mathbf{R}_{w}$ as a function of the disturbance velocity $V$ is illustrated in Fig.
\ref{fig1}. There we have assumed a pressure disturbance of Lorentzian form with a Fourier transform  $\hat P_{ext}(k)=P_
{0} e^{-bk}$, where b is the object size (set to $b=0.1\kappa^{-1}$). This choice will be taken for the figures throughout this work.
The wave resistance is equal to zero for $V < c_{min}$ and presents a discontinuous behavior at 
$V = c_{min}$ 
(see reference \cite{Chevy} and \cite{Alexei}  for a more complete discussion).

\subsection{Accelerated straight motion}

We now turn to the case where the disturbance -- initially at rest -- is suddenly set to a uniform motion 
(characterized by a constant velocity $V$) at time $t=0$.
The corresponding  trajectory  is given by $\mathbf{r}_{0}(t) \, = \; -V \, t \, \theta(t) \, \mathbf{u}_{x} $. 
As long as the perturbation does not move (i.e. for $t < 0$), the wave resistance is equal to zero.
In order to calculate the wave resistance for $t > 0$, we solve Eq.(\ref{surface2}) along with the 
initial conditions $\hat \zeta (\mathbf{k},t = 0)=0$ and $\dfrac{\partial \hat \zeta (\mathbf{k},t  = 0)}
{\partial t}=0$, yielding

\begin{equation}\label{verticaldisplacement} 
 \hat \zeta (\mathbf{k},t)= - \displaystyle{\int^{t}_{0}} \mathrm{d}\tau \dfrac{k\, \hat{P}_{ext}\left( k
\right) }{\omega \left( k\right)  \rho} e^{-i\mathbf{k}\mathbf{r}_{0}\left( \tau\right) }\sin\left(\omega \left
( k\right) \left( t-\tau \right)  \right).
\end{equation} 
Equation (\ref{Hav}) then leads to the following expression for the wave resistance

\begin{figure}
\includegraphics[width=3.5in,angle=0]{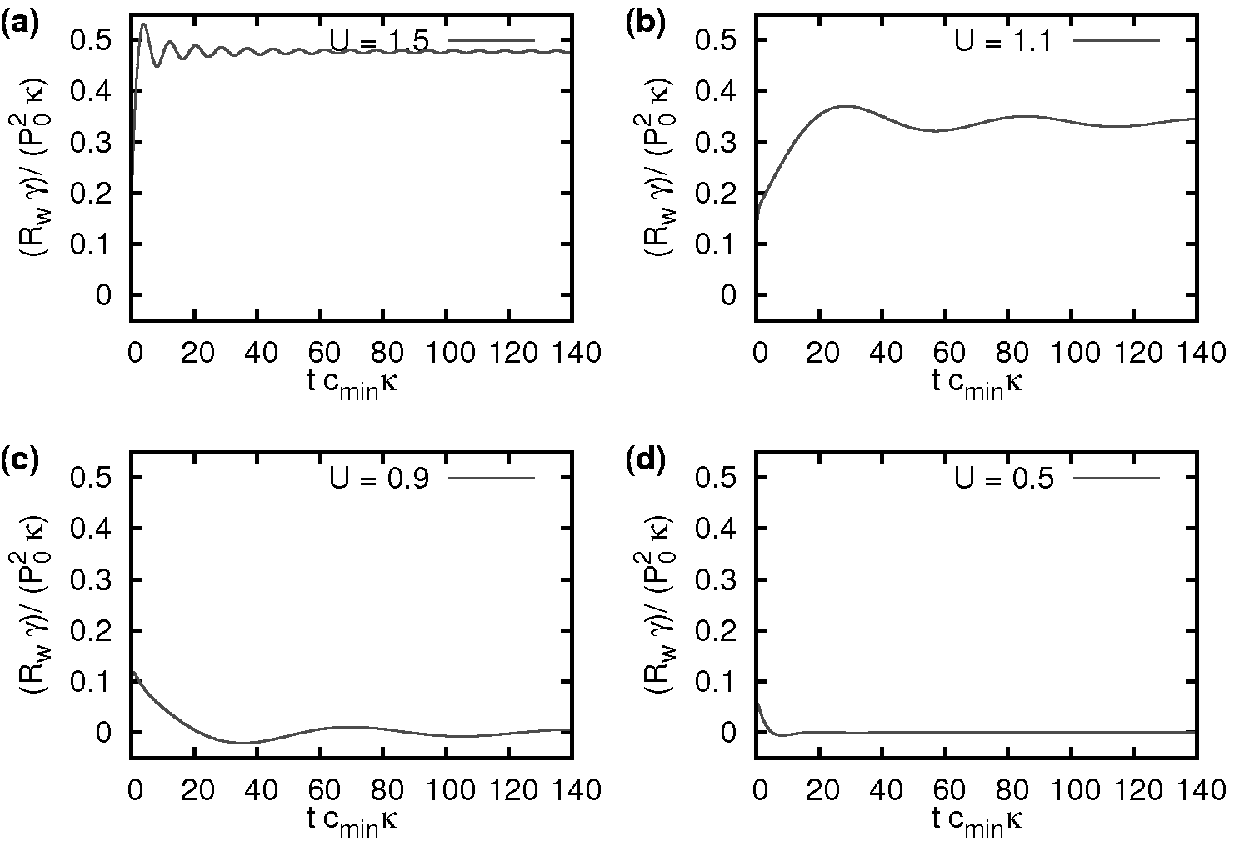}
\caption{The wave resistance $\mathrm{R}_{w}$ (in units of $P_{0}^{2}\kappa/\gamma$) is shown 
as a function of the reduced time $c_{min}\kappa t$ for an accelerated motion with different $U=V/c_
{min}$ (see Eq.(\ref{Rwaccelerated})). Respectively, panels (a), (b), (c) and (d) correspond to a 
reduced velocity $U=1.5$, $1.1$, $0.9$ and $0.5$.
}
\label{fig2}
\end{figure}

\begin{multline}\label{Rwaccelerated} 
\mathbf{R}_{w}(t)= \\ \displaystyle\int^{\infty}_{0}\!\! \dfrac{\mathrm{d}k}{2\pi} \int_{0}^{t}\!\!\mathrm{d}
u \dfrac{k^{3}|\hat{P}_{ext}(k)|^{2}}{\rho\omega(k)} \sin\left(\omega \left( k \right) u \right)J_{1}\left
( kVu\right) \, \mathbf{u}_{x},
\end{multline} 
where $J_{1}(x)$ is the first Bessel functions of the first kind.
In the long time limit, one has \cite{Grads}

\begin{equation}
\lim_{t \to \infty} \!\! \int_{0}^{t} \!\!\! \sin\!\left(\omega \left(k\right)u \right)J_{1}\!\left(kVu\right)\mathrm
{d}u 
= \!\frac{\omega(k) \theta(V\!-c(k))}{V^{2}k^{2}\sqrt{1\!-(c(k)/V)^{2}}} 
\end{equation}
and therefore the wave resistance Eq. (\ref{Rwaccelerated}) converges to the uniform straight 
motion result,
Eq. (\ref{Rwave}). The behavior of $\mathbf{R}_{w}(t)$ is represented on Fig.\ref{fig2}
for different values of the disturbance velocity and will be discussed in detail in section \ref{results}.

\subsection{Decelerated straight motion}

\begin{figure}
\includegraphics[width=3.5in,angle=0]{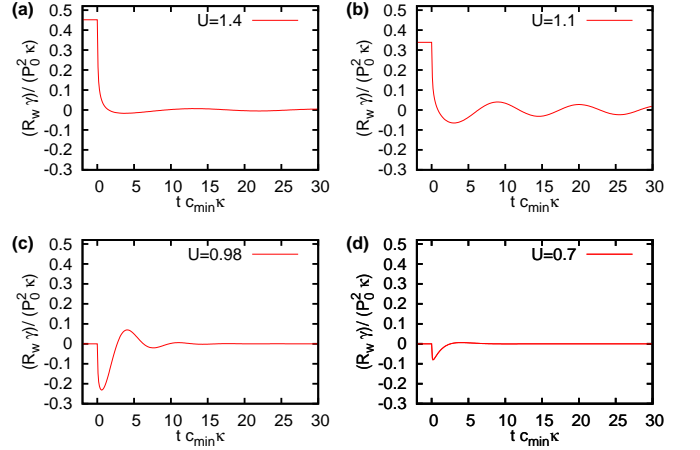}
\caption{The wave resistance $\mathrm{R}_{w}$ (in units of $P_{0}^{2}\kappa/\gamma$) is shown 
as a function of the reduced time $c_{min}\kappa t$ for a decelerated motion with different reduced 
velocities $U=V/c_{min}$ (see Eq.(\ref{Rwdecelerated})). Respectively, panels (a), (b), (c) and (d) 
correspond to a reduced velocity $U=1.4$, $1.1$, $0.98$ and $0.7$.
}
\label{fig3}
\end{figure}

Let us now consider the case of a disturbance moving with a constant velocity $V$ for $t < 0$, and
suddenly set and maintained at rest for $t > 0$. This corresponds to the following trajectory: 
$\mathbf{r}_{0}(t) \; = \; -V \, t \, \theta(-t) \,  \mathbf{u}_{x}$.
As long as $t < 0$, the wave resistance $\mathbf{R}_{w}$ is given by the uniform straight motion 
result Eq. (\ref{Rwave}).
In order to calculate the wave resistance for $t > 0$, we solve Eq.(\ref{surface2}) along with the 
initial conditions

\begin{equation}\label{continuity1} 
 \hat \zeta (\mathbf{k},t=0) \; = \; \frac{-k \hat{P}_{ext}(k)}{\rho (\omega(k)^{2}-(V k_{x})^{2})}
\end{equation} 
and
\begin{equation}\label{continuity2} 
\dfrac{\partial \hat \zeta }{\partial t}(\mathbf{k},t=0) \; = \; \frac{-k \hat{P}_{ext}(k) (i Vk_{x})}{\rho 
(\omega(k)^{2}-(V k_{x})^{2})}
\end{equation}
(where we have used Eq. (\ref{uniform})).
This leads to
\begin{multline}
\displaystyle \hat \zeta (\mathbf{k},t \geqslant 0)=
\\ \left( \!\frac{k \hat{P}_{ext}(k)}{\rho \omega(k)^{2}}\!-\!\frac{k \,\hat{P}_{ext}(k)}{\rho (\omega(k)^{2}-
(V k_{x})^{2})}\!\right) \cos\left( \omega\left( k\right) t\right) 
\\-\frac{k \,(i Vk_{x}) \,\hat{P}_{ext}(k)}{\rho (\omega(k)^{2}-(V k_{x})^{2})}\dfrac{\sin \left( \omega\left
( k\right) t\right)}{\omega(k)} -\frac{k \hat{P}_{ext}(k)}{\rho \omega(k)^{2}}.
\end{multline} 
Equation (\ref{Hav}) then leads to the following expression for the wave resistance:
\begin{multline}\label{Rwdecelerated} 
\displaystyle \mathbf{R}_{w}(t)= \\ \int_{0}^{\infty} \dfrac{\mathrm{d}k}{2\pi} \dfrac{k \rvert \hat{P}_{ext}
\left( k\right) \rvert^{2}}{\rho.V^{2}}   \cos\left( \omega \left( k\right) t \right) \dfrac{\theta \left( V-c(k)
\right) }{\sqrt{1-\left( \dfrac{c(k)}{V}\right) ^{2}}} \mathbf{u}_{x} \\ 
+\int_{0}^{\infty}\!\! \dfrac{\mathrm{d}k}{2\pi} \dfrac{k \rvert \hat{P}_{ext}\left( k\right) \rvert^{2}}{\rho.V^
{2}}   \sin\left( \omega \left( k\right) t \right)\! \left(\!\!\! \dfrac{V}{c(k)}\!-\!\dfrac{\theta \left(c(k)-V\right) }
{\sqrt{\!\left( \!\dfrac{c(k)}{V}\!\right) ^{2}\!\!\!\!-\!1}}\!\!\right)  \mathbf{u}_{x}.
\end{multline}
In the long time limit ($t \to \infty$), the Riemann-Lebesgue lemma \cite{Lebesgue}, for a Lebesgue 
integrable function $f$

\begin{equation}
\displaystyle \lim_{t \to \infty} \int f(x) \, e^{\mathit{i} \, x \, t} \mathrm{d}x \, = \, 0
\end{equation}
permits to determine the limit: the wave resistance given by Eq. (\ref{Rwdecelerated}) 
converges to $0$ for $t \to \infty$, as expected.
The behavior of $\mathbf{R}_{w}(t)$ is represented in Fig.\ref{fig3}
for different values of the disturbance velocity and will be discussed in detail in section the next 
section (Sec.\ref{results}).

\section{Results and discussion}\label{results}

\subsection{Accelerated straight motion} 

\begin{figure}
\includegraphics[width=3.5in,angle=0]{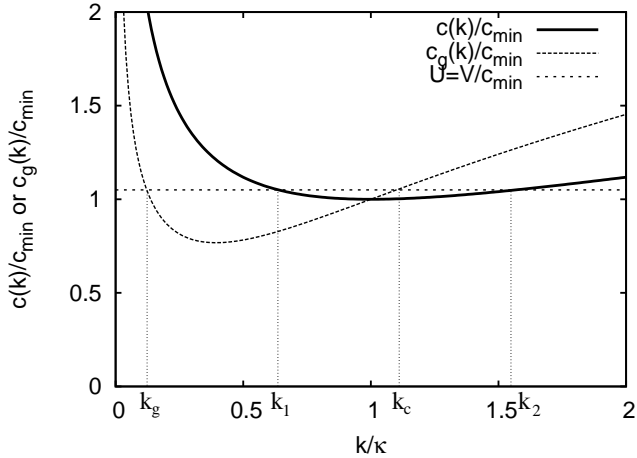}
\caption{Graphical representation of wavenumbers $k_{1}$, $k_{2}$, $k_{c}$ and $k_{g}$  in units 
of $\kappa$.  $k_{1}$ and $k_{2}$ are the solutions of the equation $c(k)=V$ and correspond to the 
intersection between the curve $c(k)/c_{min}$ and the line $U=V/c_{min}$. $k_{g}$ and $k_{c}$ are 
the solutions of the equation $c_{g}(k)=V$ and correspond to the intersection between the curve 
$c_{g}(k)/c_{min}$ and the line $U=V/c_{min}$. Analytical expresions of $k_{c}$ and $k_{g}$ can 
be given but are rather lengthy.
}
\label{fig4}
\end{figure}

Figures \ref{fig2} and \ref{fig3} represent the behavior of the wave resistance for the accelerated 
and the decelerated cases, respectively. In order to get a better understanding of 
the behavior of 
$\mathrm{R}_{w} \, = \, \mathbf{R}_{w} \cdot \mathbf{u}_{x}$,  
we will perform analytic expansions of Eq.(\ref{Rwaccelerated}) and Eq.(\ref{Rwdecelerated}), 
respectively.
Let us start with the accelerated case, Eq.(\ref{Rwaccelerated}). Since one has the product of two 
oscillating functions, $\sin\left(\omega \left( k \right) u \right)$ and $J_{1}\left( kVu\right)$, one can 
use a stationary phase approximation \cite{Riley}. The sine function oscillates with a phase $\phi_
{1}=\omega(k) u$ whereas the Bessel function $J_{1}$ oscillates with a phase $\phi_{2}=V k u$.  
Their product has thus two oscillating terms, one with a phase $\phi_{-}=\phi_{1}-\phi_{2}$ and the 
other with a phase $\phi_{+}=\phi_{1}+\phi_{2}$. According to the stationary phase approximation, 
the important wavenumbers are given by  $\dfrac{\mathrm{d}\phi_{-}}{\mathrm{d}k}=0$ and $\dfrac
{\mathrm{d}\phi_{+}}{\mathrm{d}k}=0$. The latter equation does not admit any real solution and the 
corresponding contribution to the wave resistance decreases exponentially and can be neglected.  
The equation $\dfrac{\mathrm{d}\phi_{-}}{\mathrm{d}k}=0$ leads to $(c_{g}(k) - V) u = 0$, where $c_
{g}(k) \, = \, \frac{\mathrm{d} \omega (k)}{\mathrm{d} k}$ is the group velocity of the capillary-gravity 
waves \cite{Lighthill}. This equation has solutions only if $V\geqslant \mathrm{min}(c_{g}(k))=(3\sqrt
{3}/2-9/4)^{1/4}c_{min}\approx 0.77 c_{min}$. When this condition is satisfied, two wavenumbers 
are selected: $k_{g}$ (mainly dominated by gravity) and $k_{c}$ (mainly dominated by capillary 
forces), with $k_{g}<k_{c}$ (see Fig.\ref{fig4}). If these two wavenumbers are sufficiently separated 
(that is, for velocities not too close to $0.77 c_{min}$), one then finds that  in the long time limit $
\mathrm{R}_{w}$ oscillates around its final value as

\begin{multline}\label{approxaccelerate} 
 \mathrm{R}_{w}(t)= \mathrm{R}_{w}(\infty)+\dfrac{1}{2\pi \rho \sqrt{V}}\dfrac{k_{c}^{5/2} \rvert \hat P_
{ext}(k_{c}) \rvert^{2}}{\omega(k_{c})\sqrt{\rvert\dfrac{d^{2}\omega}{\mathrm{d}k^{2}}\left(k_{c} \right) 
\rvert}}\dfrac{\cos(\Omega_{c}t)}{\Omega_{c}t} \\ + \dfrac{1}{2\pi \rho \sqrt{V}}\dfrac{k_{g}^{5/2} \rvert 
\hat P_{ext}(k_{g}) \rvert^{2}}{\omega(k_{g})\sqrt{\rvert\dfrac{d^{2}\omega}{\mathrm{d}k^{2}}\left(k_
{g} \right) \rvert}}\dfrac{\sin(\Omega_{g}t)}{\Omega_{g}t},
\end{multline} 
where $\mathrm{R}_{w}(\infty)$ is given by Eq.(\ref{Rwave}) (and is equal to zero if $V < c_{min}$),
$\Omega_{c}=(c(k_{c})-c_{g}(k_{c}))k_{c}$ and $\Omega_{g}=(c(k_{g})-c_{g}(k_{g}))k_{g}$.

Therefore, even if the disturbance velocity $V$ is smaller than $c_{min}$, there exists a transient 
nonzero wave resistance \cite{capillary} decreasing as $1/t$  (for $V>0.77 \, c_{min}$).

\begin{figure}
\includegraphics[width=3.5in,angle=0]{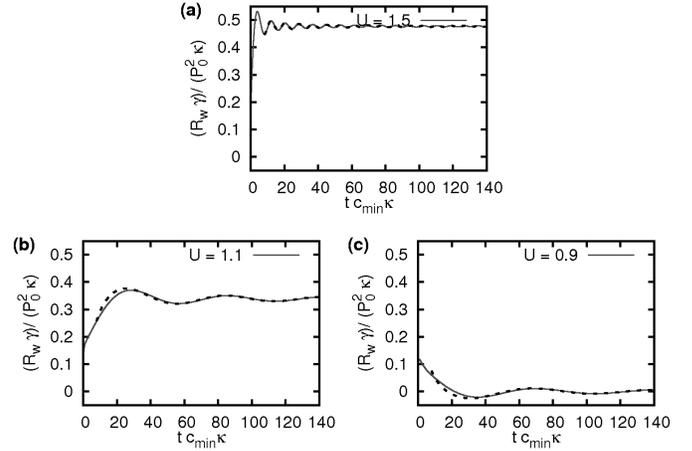}
\caption{The wave resistance $\mathrm{R}_{w}$ (in units of $P_{0}^{2}\kappa/\gamma$) is shown 
as a function of the reduced time $c_{min}\kappa t$ for an accelerated motion with different $U=V/c_
{min}$ (see Eq.(\ref{Rwaccelerated})). Respectively, panels (a), (b) and (c) correspond to a 
reduced velocity $U=1.5$, $1.1$ and $0.9$. The solid lines are obtained by a numerical integration 
of Eq.(\ref{Rwaccelerated}). The dashed lines correspond to the asymptotic expansion given by Eq.
(\ref{approxaccelerate}).
}
\label{fig5}
\end{figure}

We obtain a good agreement between the numerical calculation of Eq.(\ref{Rwaccelerated}) and 
the analytical approximation Eq.(\ref{approxaccelerate}) as shown on Fig.\ref{fig5}. Note that the oscillations displayed by the wave resistance are characterized by a period $2 \pi/
\Omega_{c}$ or $2 \pi/\Omega_{g}$ that diverges as $V$ approaches $c_{min}$. In the particular 
case where $V  \gg c_{min}$, Eq.(\ref{approxaccelerate}) reduces to 

\begin{multline}\label{approxacceleratelargeV} 
 \mathrm{R}_{w}(t) \; = \; \mathrm{R}_{w}(\infty) \, + \, \dfrac{1}{2^{9/2} \pi} \dfrac{\kappa^{} c_{min}^2 
\rvert \hat P_{ext}(k_{g}) \rvert^{2}}{\rho V^5 t} {\sin\left(\! \frac{c_{min}^2 \kappa t}{8 V}\!\right) } \\ - \, 
\dfrac{2^{3/2}}{ \pi} \dfrac{\kappa^{}  \rvert \hat P_{ext}(k_{c}) \rvert^{2}}{c_{min}^2 \rho V t} {\cos\left
( \!\frac{8 V^3 \kappa t}{27 c_{min}^2}\!\right) }.
\end{multline} 
Let us  now give a physical interpretation for the wave resistance behavior as described by Eq.(\ref
{approxaccelerate}): 
during the sudden acceleration of the disturbance (taking place at $t = 0$), a large range of
wavenumbers is emitted. Waves with wavenumbers $k$ such that $c_{g}(k)>V$ 
will move faster than the disturbance (which moves with the velocity $V$),
whereas waves with wavenumbers $k$ such that $c_{g}(k) < V$ 
will move slower. The main interaction with the moving disturbance will therefore correspond 
to wavenumbers satisfying $c_{g}(k) = V$, that is to $k_c$ and $k_g$ (hence their appearance in
Eq.(\ref{approxaccelerate})). Due to the Doppler effect, the wave resistance $\mathrm{R}_{w}$ 
(which is the force
exerted by the fluid on the moving disturbance) 
oscillates with an angular frequency $(c(k) - V) k$, hence the appearance of $\Omega_{c}$ and $
\Omega_{g}$
in Eq.(\ref{approxaccelerate}).

Note that in the case $V < 0.77 c_{min}$, the wave resistance, Eq.(\ref{Rwaccelerated}), is nonzero 
and
decreases exponentially with time (see Fig.\ref{fig2}(d)).

\begin{figure}
\includegraphics[width=3.3in,angle=0]{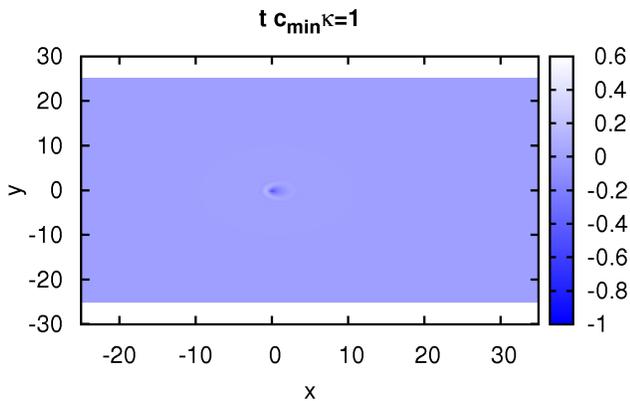}
\caption{Accelerated straight motion: Transient vertical displacement (in units of $P_{0}\kappa/(\rho 
c_{min}^{2})$) of the free surface at $t c_{min} \kappa = 1$ obtained by inverse Fourier transform 
of Eq.(\ref{verticaldisplacement}).
Note that the surface disturbance is localized around the object and close to the one obtained by a stone thrown in water.}
\label{fig6}
\end{figure}

\begin{figure}
\includegraphics[width=3.3in,angle=0]{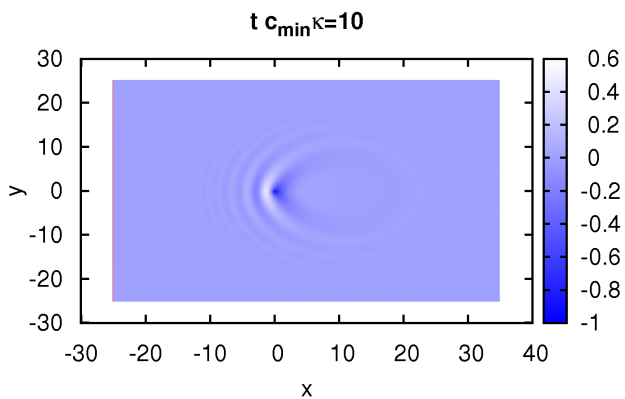}
\caption{Accelerated straight motion: Transient vertical displacement (in units of $P_{0}\kappa/(\rho 
c_{min}^{2})$) of the free surface at $t c_{min} \kappa = 10$ obtained by inverse Fourier transform 
of Eq.(\ref{verticaldisplacement}). One can already see ahead of the disturbance the waves 
associated with $k_{1}$, while the ones associated to $k_{2}$ are less well formed.
}
\label{fig7}
\end{figure}

\begin{figure}
\includegraphics[width=3.3in,angle=0]{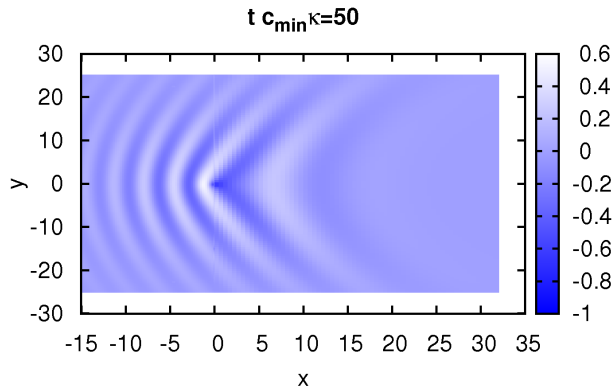}
\caption{Accelerated straight motion: Transient vertical displacement (in units of $P_{0}\kappa/(\rho 
c_{min}^{2})$) of the free surface at $t c_{min} \kappa = 50$ obtained by  inverse Fourier transform 
of Eq.(\ref{verticaldisplacement}).
}
\label{fig8}
\end{figure}

In order to get a better physical picture of the generated wave patterns, we have 
also calculated numerically the transient vertical displacement of the free surface  $\zeta(\mathbf
{r},t)$ in the accelerated case (Eq.(\ref{verticaldisplacement})). The corresponding patterns (as 
seen in the frame of the moving object), are presented on Figs. \ref{fig6}, \ref{fig7} and \ref{fig8} for 
different reduced times $c_{min} \kappa t$ ($1$,\, $10$ and $50$, respectively). At $c_{min} \kappa 
t =1$, the perturbation of the free surface is very localized around the disturbance (close to the 
same one obtained by a stone's throw). At $c_{min} \kappa t =10$, some capillary waves can 
already be observed at the front of the disturbance. At $c_{min} \kappa t =50$, on can see a  V-shaped pattern that prefigure the steady pattern of the uniform straight motion (obtained at very 
long times). These predictions concerning the wave pattern might be compared to experimental 
data using the recent technique of Moisy, Rabaud, and Salsac \cite{Moisy}.

\subsection{Decelerated  straight motion}
Let us now turn to the case of the decelerated motion, described by Eq.(\ref{Rwdecelerated}). We 
first will assume that 
$V > c_{min}$. In that case, the denominators appearing in the integrals vanish for $k$ such that $ 
c(k) = V$, 
that is for $k_{1}=\kappa (U^{2}-\sqrt{U^{4}-1})$ (mainly dominated by gravity) and $k_{2}=\kappa 
(U^{2}+\sqrt{U^{4}-1}))$ (mainly dominated by capillary forces) as shown in Fig.\ref{fig4}, where 
$U=V/c_{min}$.
Both wavenumbers contribute to the wave resistance. Using the stationary phase approximation 
one then finds that  in the long time limit $\mathrm{R}_{w}$ oscillates as $\cos\left(Vk_{1}t+ \pi/
4\right)/\sqrt{t}$. 
More precisely, one has (for large $t$)

\begin{equation}\label{fitdece} 
R_{w}(t)=\dfrac{1}{\sqrt{\pi}\,V\,\sqrt{\rho \gamma}} \dfrac{k_{1}^{3/2} \rvert \hat P_{ext}(k_{1}) \rvert^
{2}}
{ \sqrt{ \,(k_{2}-k_{1})\,c_{g}(k_{1}) \,t}}\cos(Vk_{1}t+ \pi/4).
\end{equation} 

\begin{figure} 
\includegraphics[width=3.3in,angle=0]{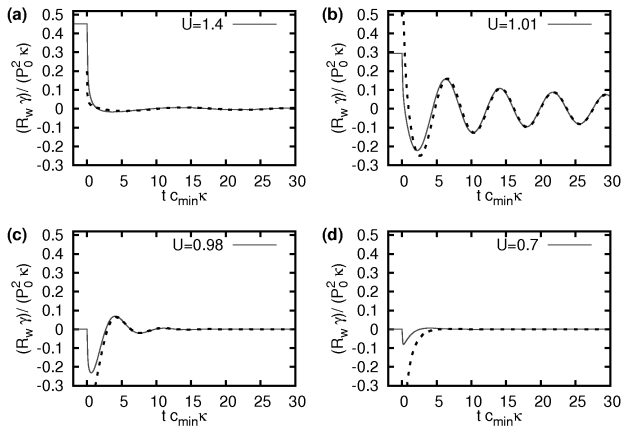}
\caption{The wave resistance $\mathrm{R}_{w}$ (in units of $P_{0}^{2}\kappa/\gamma$) is shown 
as a function of the reduced time $c_{min}\kappa t$ for a decelerated motion with different reduced 
velocities $U=V/c_{min}$ (see Eq.(\ref{Rwdecelerated})). Respectively, panels (a), (b), (c) and (d) 
correspond to a reduced velocity $U=1.5$, $1.01$, $0.98$ and $0.7$. The dashed lines 
correspond to the asymptotic expansion given by Eq.(\ref{fitdece}).
}
\label{fig9}
\end{figure}

The wave resistance displays oscillations that are characterized by a period $2 \pi/(Vk_{1})$.
Figure \ref{fig9} shows the good agreement between the numerical calculation of Eq.(\ref
{Rwdecelerated}) and the analytical approximation Eq.(\ref{fitdece}) for large times ($c_{min} \kappa 
t >10$).
Again one can give a simple physical interpretation of the wave resistance, Eq.(\ref{fitdece}). The 
period of the oscillations
  is given by $2 \pi /(V k_{1})$ and depends only on the wavenumber $k_{1}$ (mainly dominated by 
gravity), even 
  in the case of an object with a size $b$ much smaller than the capillary length. This can be 
understood as follows. 
  For $t < 0$, the disturbance moves with a constant velocity $V$ and emits waves with 
wavenumber
  $k_{1}$ and  waves with wavenumber $k_{2}$ (mainly dominated by capillary forces). The $k_
{1}$-waves lag behind the disturbance, while the $k_{2}$-waves move ahead of it. When the 
disturbance stops at $t = 0$, the $k_{2}$-waves keep moving forward and do not interact 
  with the disturbance. However, the $k_{1}$-waves
  will encounter the disturbance and interact with it. Hence the period $2 \pi /(V k_{1})$ of the wave 
resistance
  Eq.(\ref{fitdece}). 
  The $1/\sqrt{t}$ decrease of the magnitude of wave resistance in Eq.(\ref{fitdece}) can be
  understood as follows.  At time $t > 0$, the disturbance (which is at rest at $x = 0$) is hit by a $k_
{1}$-wave 
  that has previously been emitted distance $c_{g}(k_{1})t$ away from it. The vertical amplitude $
\zeta$
  of this wave is inversely proportional to the square root of this distance. Indeed, as the liquid is 
inviscid, the
  energy has to be conserved and one thus has $\zeta \propto 1/\sqrt{c_{g}(k_{1})t}$.
  Since the wave resistance $\mathrm{R}_{w}$ is proportional to vertical amplitude of the wave $
\zeta$,
  on recovers that $\mathrm{R}_{w}$ decreases with time as $1/\sqrt{c_{g}(k_{1})t}$.

In the particular case where $V  \gg c_{min}$, Eq.(\ref{fitdece}) reduces to:

\begin{equation}\label{fitdeceVlarge} 
R_{w}(t) = \dfrac{1}{2 \sqrt{\pi}} \dfrac{\kappa^{3/2}\,\rvert \hat P_{ext}(k_{c}) \rvert^{2}\, c_{min}^3}
{ \rho \, V^{11/2}\, \sqrt{t}} \cos\left( \dfrac{\kappa c_{min}^{2}}{2 V} t+ \pi/4\right).
\end{equation}

Let us now discuss the case $V < c_{min}$ (still considering the decelerated motion Eq.(\ref
{Rwdecelerated})). In that case, the wave resistance oscillates (in the long time limit)
approximatively as $\left(e^{-U \sqrt{1-U^{4}} t \kappa c_{min}}/\sqrt{t \kappa c_{min}}\right) \sin(U^
{3} t \kappa c_{min}+\chi)$.
More precisely, one has (for large $t$) 

\begin{multline}\label{fitdece2} 
\mathrm{R}_{w}(t)=\sqrt{\dfrac{2}{\pi}}\dfrac{\kappa^{2}\rvert \hat P_{ext}(\tilde k) \rvert^{2}}{\rho c_
{min}^{2}}\dfrac{\sin(U^{3} \, c_{min} \, \kappa \, t+\chi)}{\sqrt{ \kappa \, c_{min}\, t }} \\ \dfrac{  e^{-U 
\sqrt{1-U^{4}} t \kappa c_{min}}}{\sqrt{2} \pi U^{1/2} (1-U^{4})^{1/4}(3U^{4}+1)^{1/4}},
\end{multline} 
where 
\begin{multline}
\chi= (3/2) \arctan \left( \sqrt{1-U^{4}}/U^{2}\right) \\
  - (1/2) \arctan \left( \sqrt{1-U^{4}}/(2U^{2})\right)
\end{multline} 
and $\tilde k = \kappa \left( U^{2} + i \sqrt{1-U^{4}}\right) $. Note that in the case $V < c_{min}$, the 
wave resistance Eq.(\ref{fitdece2}) also displays some oscillations, although no waves
were emitted at $t < 0$. These oscillations - that have a exponential decay in time - might be due to 
the sudden arrest  of the disturbance. Such oscillations could also be present in the case  $V > c_
{min}$ but
are hidden by the interaction between the $k_{1}$-waves and the disturbance which lead to a 
much more slower 
$1/\sqrt{t}$ decay.

\section{Conclusions}\label{conclusions} 

In this article, we have shown that a disturbance undergoing a rectilinear accelerated or 
decelerated motion at a liquid-air interface emits waves even if its velocity $V$ (the final one in the 
accelerated case and the initial one in the decelerated case, respectively) is smaller than $c_{min}
$. This corroborates the results of Ref.\cite{Alexei}. For this purpose, we treat the wave emission problem 
by a linearized theory in a Monge representation. Then, we derive the analytical expression of the wave 
resistance and solve it by numerical integration. Asymptotic expansions permit to extract the predominant behavior of the wave 
resistance. Some vertical displacement patterns are also calculated in order to shown how the waves invade 
the free surface.

The results presented in this paper should be important for a better understanding of the propulsion 
of water-walking insects \citep{bush,deny,McNeill,Buhler}, like whirligig beetles, where accelerated 
and decelerated motions frequently occurs (e.g., when hunting a prey or escaping a predator \cite
{Voise}). Even in the case where the insect motion appears as rectilinear and uniform, one has to 
keep in mind that the rapid leg strokes are accelerated and might produce a wave drag even below 
$c_{min}$. The predictions concerning the wave patterns might be compared to experimental 
data using the recent technique of Moisy, Rabaud, and Salsac \cite{Moisy}.

It will be interesting to take in our model some non-linear effects \cite{Dias} because the waves radiated by whirligig beetles \cite{Voise} have a large amplitude.
Recently Chepelianskii \textit{et al.}  derived a self-consistent integral equation describing the flow 
velocity around the moving disturbance \cite{mickael}. It would be interesting to incorporate this 
approach into the present study.

\begin{acknowledgements}
We thank Jonathan Voice and Fr\'{e}d\'{e}ric Chevy for useful discussions and Falko Ziebert for a 
critical reading of the manuscript.
\end{acknowledgements}


\begin{thebibliography}{10}
\bibitem{Fourier}
The Fourier transform $\hat f(\mathbf{k},t)$ is related with the function $f(\mathbf{r},t)$ through $f
(\mathbf{r},t)=\int \frac{\mathrm{d}^2 k}{(2\pi)^{2}} e^{i \mathbf{k}. \mathbf{r}}\hat f(\mathbf{k},t)$
\bibitem{capillary}

In the particular case  of a Lorentzian pressure disturbance with a Fourier transform  $\hat P_{ext}
(k)=P_{0} e^{-bk}$, where b is the object size (set to $b=0.1\kappa^{-1}$), we have  numerically 
checked that up to  $V \approx 8 c_{min}$
the main contribution to the wave resistance is due to $k_{c}$.
\end{thebibliography}
\end{document}